\documentclass[journal]{IEEEtran}
\ifCLASSINFOpdf
\else
\fi

\usepackage{tikz}
\usepackage{amsmath}
\usepackage{subfigure}
\usepackage{amsthm}

\usepackage{tabularx}
\usepackage{pifont}
\usepackage{filecontents}
\usepackage{amsmath,amssymb,amsfonts}
\usepackage{url}
\usepackage{caption}
\usepackage{colortbl}
\newcommand{\circled}[1]{\normalsize{\textcircled{\scriptsize{#1}}}\normalsize\;}

\definecolor{seagreen}{rgb}{0.18, 0.55, 0.34}
\definecolor{royalpurple}{rgb}{0.47,0.32,0.66}
\definecolor{brown(traditional)}{rgb}{0.59, 0.29, 0.0}
\definecolor{blue}{rgb}{0.3, 0.2, 0.9}

\begin{document}
%
% Efficient
% \title{Towards Efficient Wireless Network Management Leveraging Deep Generative Models: A Tutorial and Case Study} 
\title{Deep Generative Model and Its Applications in Efficient Wireless Network Management: A Tutorial and Case Study}

\author{Yinqiu~Liu,
        Hongyang~Du,
        Dusit~Niyato,~\IEEEmembership{Fellow,~IEEE},
        Jiawen~Kang,
        Zehui~Xiong,
        Dong In Kim,~\IEEEmembership{Fellow, IEEE},
        and Abbas Jamalipour,~\IEEEmembership{Fellow, IEEE}
        \thanks{Y. Liu, H. Du, and D. Niyato are with the School of Computer Science and Engineering, Nanyang Technological University, Singapore (e-mail: yinqiu001@e.ntu.edu.sg, hongyang001@e.ntu.edu.sg, and dniyato@ntu.edu.sg).}
        \thanks{J. Kang is with the School of Automation, Guangdong University of Technology, China (e-mail: kavinkang@gdut.edu.cn).}
        \thanks{Z. Xiong is with the Pillar of Information Systems Technology and Design, Singapore University of Technology and Design, Singapore (e-mail: zehui xiong@sutd.edu.sg).}
        \thanks{D. I. Kim is with the Department of Electrical and Computer Engineering, Sungkyunkwan University, South Korea (e-mail: dikim@skku.ac.kr).}
        \thanks{A. Jamalipour is with the School of Electrical and Information Engineering, University of Sydney, Australia (e-mail: a.jamalipour@ieee.org).}% <-this % stops a space
        \thanks{Corresponding author: D. I. Kim.}
        \vspace{-1.4em}}% <-this % stops a space
\maketitle
\begin{abstract}
With the phenomenal success of diffusion models and ChatGPT, deep generation models (DGMs) have been experiencing explosive growth from 2022.
Not limited to content generation, DGMs are also widely adopted in Internet of Things, Metaverse, and digital twin, due to their outstanding ability to represent complex patterns and generate plausible samples.
In this article, we explore the applications of DGMs in a crucial task, i.e., improving the efficiency of wireless network management.
Specifically, we firstly overview the generative AI, as well as three representative DGMs.
Then, a DGM-empowered framework for wireless network management is proposed, in which we elaborate the issues of the conventional network management approaches, why DGMs can address them efficiently, and the step-by-step workflow for applying DGMs in managing wireless networks.
Moreover, we conduct a case study on network economics, using the state-of-the-art DGM model, i.e., diffusion model, to generate effective contracts for incentivizing the mobile AI-Generated Content (AIGC) services.
Last but not least, we discuss important open directions for the further research.
\end{abstract}


% \fi
\IEEEpeerreviewmaketitle
\vspace{-0.5cm}
\section{Introduction}
Artificial Intelligence (AI) has become indispensable and ubiquitous in academic research, industrial production, and daily life.
Tailoring to various tasks, researchers have proposed a series of AI models using diverse architectures (e.g., artificial neural networks and support vector machines) and learning strategies (e.g., supervised learning and unsupervised learning).
From the high-level perspective, i.e., the learning objective, these models can be divided into two types, namely discriminative models and generative models.
In the past decade, we witnessed the success of discriminative AI, which generously supports classification tasks and facilitates numerous smart applications such as computer vision.

Instead of learning the boundary between classes, generative models can learn the latent representation of the given samples, then generate plausible content.
In 2022, generative AI, especially pre-trained large models, experienced explosive development.
The underlying reasons are twofold.
Firstly, the advances in hardware, e.g., GPU, significantly improve the computing power that can be invested in model training. 
Moreover, leveraging deep neural networks, generative models evolve into Deep Generative Models (DGM), with outstanding capability to represent complex patterns \cite{DGM}.
DGMs have achieved phenomenal success in the following fields.
\begin{itemize}
    \item \textbf{Image generation:} In Aug. 2022, \textit{Stable Diffusion} was published by StabilityAI. As a text-to-image generation model, Stable Diffusion allows users to generate realistic images only using several sentences of prompts.  
    \item \textbf{Chatbot:} In Nov. 2022, Open AI published \textit{Chat Generative Pre-trained Transformer} (ChatGPT) and attracted global attention. As a chatbot, ChatGPT enables users to complete diverse language tasks, including translation, text summarization, Q\,\&\,A, even coding. In Mar. 2023, ChatGPT was further upgraded using the GPT-4 model, which can process multimodal inputs (e.g., images).
    \item \textbf{Music composition:} In Jan. 2023, Google published an automatic music composition model named MusicLM. With simple text prompts defining the theme, emotion, musical instruments, and style, MusicLM can generate high-fidelity music at 24 kHz.
\end{itemize}

The above examples demonstrate the superior generation and generalization ability of DGMs.
Not limited to content generation, DGMs have also been widely adopted on the Internet of Things (IoT), Metaverse, and digital twin \cite{IoT}.
Nonetheless, as the foundation of these emerging paradigms, DGMs for managing wireless networks are rarely researched.
Currently, wireless network management mainly relies on discriminative AI, especially Deep Reinforcement Learning (DRL) models, and other non-learning approaches such as stochastic optimization and game theory. 
Nevertheless, training DRL models requires enormous real-world data, which might be unavailable if the network is idle. 
In addition, the randomness in wireless channels caused by noise and interference is not considered because the aforementioned approaches only map the given state to an ``optimal'' strategy, without exploring these latent factors.
Finally, these approaches can hardly fit the mobility of wireless devices. 
If the structure of network states (e.g., the number of connected devices and their buffer states) changes dramatically, we have to re-formulate the problem or re-train the model from scratch.
Fortunately, DGMs show great potential to solve these problems.
\begin{itemize}
    \item \textbf{Data augmentation:} DGMs can generate realistic content by representing the given samples using latent vectors and learning their distribution. Therefore, we can overcome data scarcity by synthesizing training sets.
    \item \textbf{Latent space representation:} In the above process, the latent factors existing in the wireless networks become learnable. Consequently, we can explore the effects caused by randomness and refine our strategies accordingly to improve the flexibility. 
    \item \textbf{Knowledge transfer:} DGMs can unify the distribution of newly generated data to a given one, thereby transferring the learned knowledge among domains.
\end{itemize}

In this article, we provide a systematic tutorial on DGMs, as well as their applications in wireless network management.
Specifically, we first overview the generative AI and representative DGMs. 
Then, we propose a DGM-empowered wireless network management framework.
Based on our framework, we discuss the challenges that we intend to solve, why DGMs can help, how to apply them for managing wireless networks, and three use cases.
Furthermore, we conduct a case study on network economics, focusing on designing a flexible incentive mechanism for mobile AI-Generated Content (AIGC) services via diffusion models \cite{diffusion}, the SOTA DGM. 
\textit{To the best of our knowledge, this is the first work methodically showing why and how DGMs can improve the efficiency of wireless network management.}
Our contributions are summarized as follows:
\begin{itemize}
    \item We provide a comprehensive overview of the generative AI, including technical basics, three representative DGMs, namely Variational Autoencoder (VAE), Generative Adversarial Network (GAN), and diffusion model, and their applications beyond content generation.  
    \item We present a novel framework for wireless network management leveraging DGMs. Based on the proposed framework, we show a step-by-step workflow for DGM-empowered wireless network management and three use cases, namely network routing, resource allocation, and network economics.
    \item We conduct a case study, in which we design a diffusion-based contract theory for incentivizing mobile AIGC services. Contributed to the diffusion process, the generated contracts lead to higher utility than traditional DRL.
\end{itemize}

\section{Deep Generative Models: Basics, Current Progress, and Applications}
In this section, we comprehensively overview the generative AI, including the basics, the representative DGMs, and their applications beyond content generation.

\subsection{Basics}
Firstly, we give a brief introduction to DGMs in terms of objectives and model architectures.

\subsubsection{Objectives}
As aforementioned, AI models can be classified into two categories, namely discriminative models and DGMs, according to the learning objective.
The former makes predictions on unseen data based on conditional probability, and thus can be used for various classification tasks.
In contrast, DGMs, as the name suggests, focus on synthesize plausible samples.
For example, given a set of samples $X$ and labels $Y$, the discriminative models can capture $P(Y|X)$, i.e., the probability of $X$ belonging to class $Y.$
DGMs, however, return the joint probability $P(X, Y)$, which means the probability that $X$ can be generated.

\subsubsection{Model Architectures}
As illustrated in Fig. 1, compared with discriminative models, DGMs are featured by the generators (e.g., the decoder of VAE and the denoising process of the diffusion model).
Such generators are generally implemented by deep neural networks.
Accordingly, the training of the DGMs is just the process for refining the generators' skills. 
To be specific, the generators are trained to create plausible samples $X$, whose distribution $P_G$ approaches the real sample distribution $P_{data}$.
%Different models adopt different ways to reach this goal, which are demonstrated in the following parts. 
After training, we can leverage the generators for creating new samples. 
In the following parts, we introduce three famous DGMs, namely VAE, GAN, and diffusion model.
The architecture and features of these three models are shown in Fig. 1.
\begin{figure*}[tpb]
\centering
\includegraphics[width=2\columnwidth]{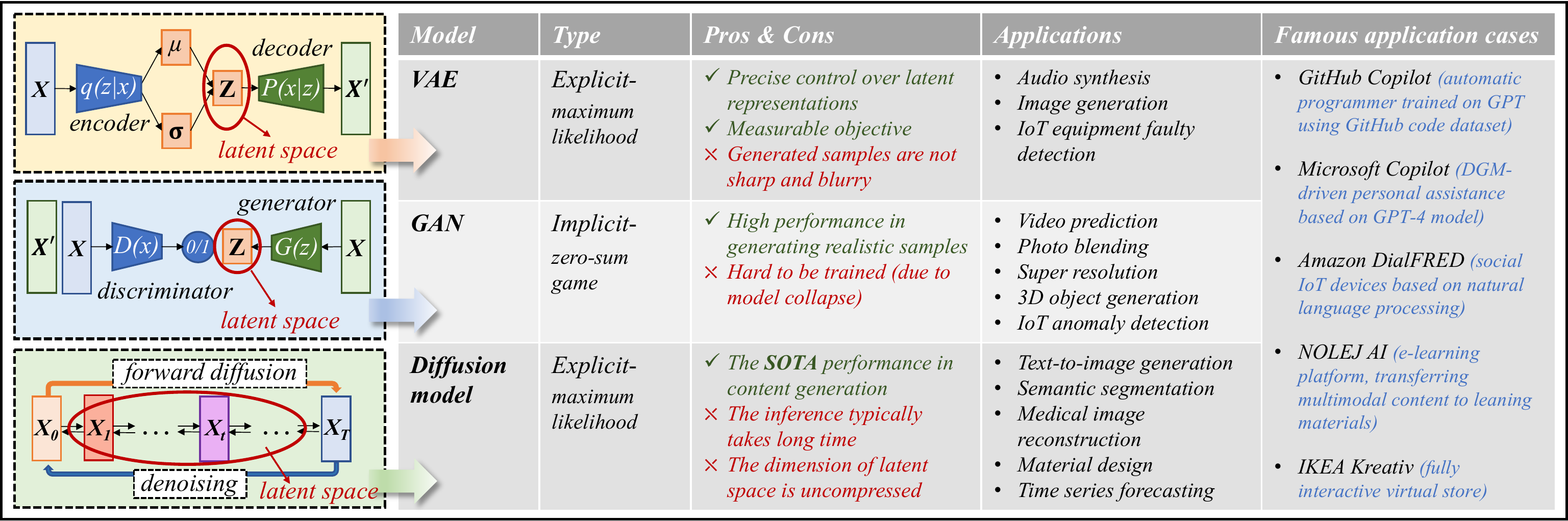}
\caption{The architectures of the mainstream DGMs and their comparisons.}
\vspace{-0.3cm}
\label{ledger architecture}
\end{figure*}
\vspace{-0.2cm}
\subsection{Variational Autoencoder (VAE)}
The idea of VAE derives from autoencoders, which firstly utilize an encoder to compress the original samples to low-dimensional latent vectors.
Then, a decoder is employed to re-construct the original input samples from latent vectors.
The higher the similarity level between the original and the reconstructed samples, the greater the model accuracy.
Obviously, although this architecture is suitable for data compression/recovery, its generation ability facing unseen samples is poor.
To improve generalization ability, VAE extends the autoencoder by mapping the input to a distribution rather than fixed vectors.
Specifically, we suppose that each input $x$ can be mapped to a latent representation $z$ with the probability $q(z|x)$, which follows a normal distribution.
Then, $z$ can be utilized to re-construct $x$, with the probability $P(x|z)$.
Therefore, VAE is trained by i) finding the distribution of $q(z|x)$, whose \textit{mean} and \textit{variance} are the parameters, ii) sampling $z$ following the distribution, and iii) reconstructing $x$ using $z$.
The corresponding loss consists of two parts, namely the divergence between $q(z|x)$ and the standard normal distribution (we assume $q(z|x)$ should follow a standard normal distribution ideally) and the difference between the original input and the generated output.
VAE is featured by the measurable loss functions and the precise control over latent representation $z$.
%It is widely adopted for image restoration, speech generation, even drug design.

\subsection{Generative Adversarial Network (GAN)}
To synthesize plausible content, we need to learn what distribution that the training set follows, denoted by $P_{data}$. 
However, given that the distributions of real-world data (e.g., images) are generally high-dimensional, representing $P_{data}$ is an intractable task. 
GAN \cite{GAN}, as a kind of implicit model, does not need the specific likelihood function to model $P_{data}$.
Instead, it employs a generator and a discriminator to conduct a zero-sum game\footnote{Zero-sum game refers to a situation, in which one party's gain is equivalent to another's loss. Hence, the overall change in benefit is zero.}.
Specifically, the generator is responsible for creating plausible samples that can fool the discriminator.
Accordingly, the discriminator is used for distinguishing the generated samples from existing samples.  
The training of GAN models follows an iterative way: i) fixing the generator and training the discriminator, to improve its distinguishing accuracy and ii) fixing the discriminator and training the generator to improve its generating skills.
After training, we can discard the discriminator and only leverage the generator to complete the generation tasks.
Compared with VAE, although GAN loses some explainability, it shows higher capability in generating high-quality multimodal contents.
%The successful applications of GAN include handwriting creation, photograph editing, video prediction, etc.

\subsection{Diffusion Model}
As shown in Fig. 1, the architecture of diffusion models is greatly different from VAE and GAN, which generate samples using only one step.
Inspired by non-equilibrium thermodynamics, diffusion models, however, consist of two Markov processes, namely forward diffusion and denoising.
Starting from the original input $X_0$, the former process gradually perturbs data by adding Gaussian noise until achieving a pure noise $X_T$. 
Given that these noise-adding steps are independent, the forward diffusion follows a normal distribution.
Then, the goal is to train the transition kernel, i.e., the probability of outputting $X_{i-1}$ given $X_i$, $\forall i \in \{1, \dots, n\}$, with which the model can reconstruct the original input from a random noise step by step.
For simplicity, the transition kernel is also assumed to follow a normal distribution, whose mean includes the parameters that need to be learned.
Therefore, a U-Net (\textit{https://en.wikipedia.org/wiki/U-Net}) is employed, using the maximum likelihood principle to learn the parameters.
After training, the model can generate plausible samples following the denoising process.
Compared with VAE and GAN, diffusion models can better match the inputs' distributions because of the fine-grained denoising processes.
Accordingly, the sample generation takes longer time.

\subsection{Applications Beyond Content Generation}
DGMs are widely adopted for creating and fine-tuning multimodel contents, e.g., text and images.
Nowadays, their applications even go beyond content generation, supporting more advanced paradigms.
In this part, we elaborate several scenarios in which DGMs also play an important role.

\subsubsection{IoT}
DGMs are applied in IoT for multiple purposes \cite{IoT}. 
For instance, given the large volumes of IoT data, detecting anomalies caused by system failure and environment change is challenging. 
To this end, we can train a DGM to learn the distribution of normal IoT data. 
Thus, the abnormal samples, which are poorly encountered, can be detected autonomously.
Similarly, such a method can also support IoT device monitoring, trust-boundary protection, etc.

\subsubsection{Metaverse}
Metaverse can be regarded as a new form of the Internet, breaking the boundary between the physical world and the virtual worlds. 
Apart from creating personalized 3D avatars, DGMs can also compress the volume of data for constructing XR environments by representing complex virtual objects using low-dimensional latent vectors.

\subsubsection{Digital Twin}
Digital twin intends to build high-fidelity virtual replicas of physical entities. 
DGMs can predict when the physical entities are likely to fail or require maintenance. This is achieved by training the DGM to learn the distributions of failures based on historical data. 

\textit{\textbf{Lessons Learned}}: The above cases motivate us to explore the applications of DGMs in wireless network management.
As the building block of these advanced paradigms, wireless networks show great functional and structural similarities with them.
Consequently, we believe DGMs can also be adopted for improving the efficiency of managing wireless networks.
In the following section, we discuss the DGM-empowered wireless network management in detail.
\begin{figure*}[tpb]
\centering
\includegraphics[width=1.7\columnwidth]{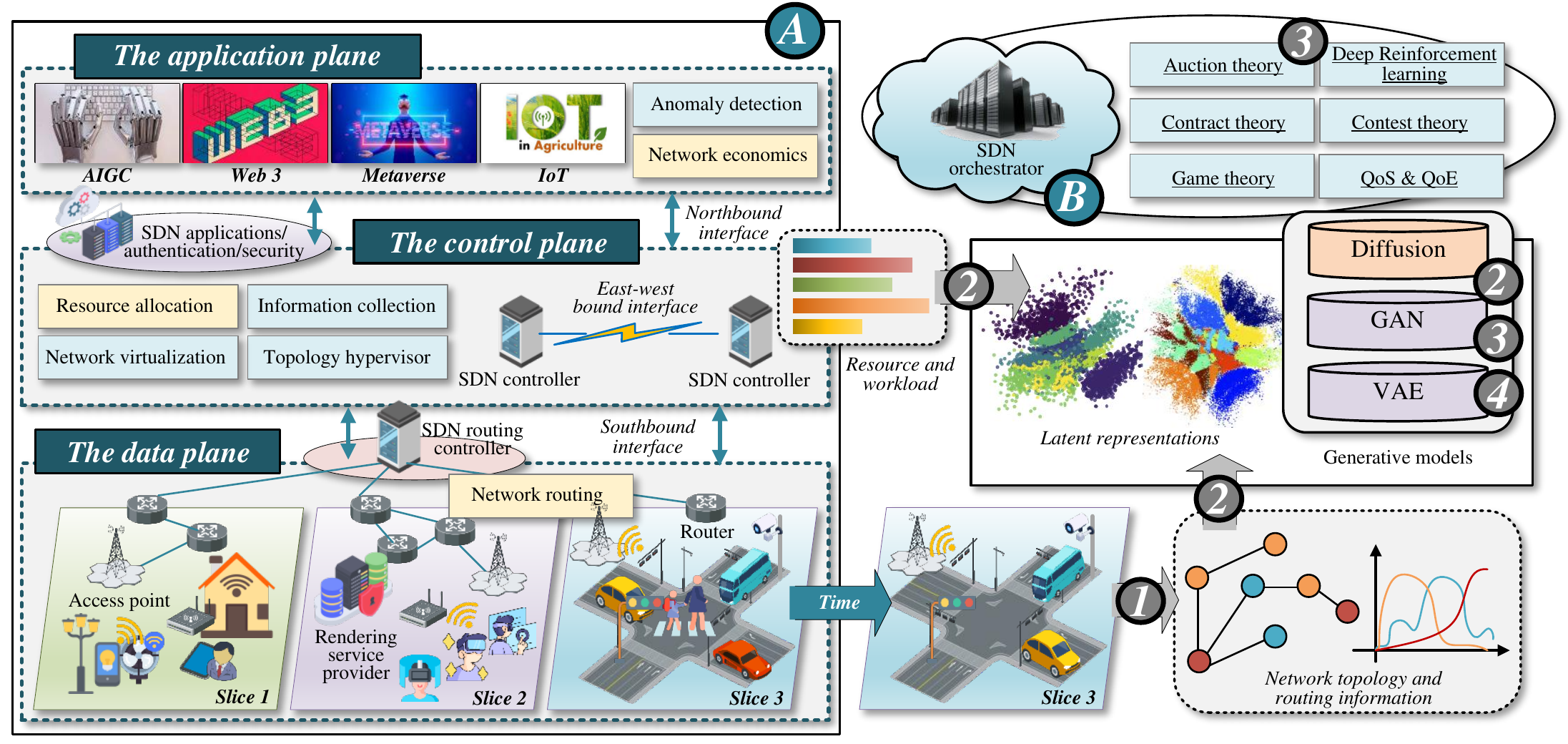}
\caption{The proposed framework. Part A: the three-layer wireless network architecture. Part B: the SDN orchestrator and its network management approaches (including DGMs).}
\vspace{-0.3cm}
\label{ledger architecture}
\end{figure*}

\section{DGMs in Efficient Wireless Network Management}
We first present a DGM-empowered framework for efficiently managing the wireless networks.
Based on the proposed framework, we answer three questions: i) what are the major issues in wireless network management, ii) why DGMs can address these issues, and iii) how to apply DGMs to manage wireless networks efficiently.

\subsection{Framework Overview}
Our framework is illustrated in Fig. 2.
Specifically, it follows a layered architecture and consists of three planes, i.e., the data plane, the control plane, and the application plane.
\begin{itemize}
    \item \textbf{The data plane}: This plane accommodates numerous wireless devices (e.g., mobile phones, smart sensors, VR headsets, and vehicles). By Software-Defined Networking (SDN) protocols like OpenFlow \cite{SDN}, the entire plane is virtualized into three slices, each of which tailors to fulfill diverse demands requested by a particular application.
    \item \textbf{The control plane}: Containing various SDN controllers, the control plane acts as the software-based coordinator of the entire framework. Note that to improve the efficiency of managing wireless networks, our framework equips SDN controllers with a module containing various DGMs, such as VAE, GAN, and diffusion model. More details are elaborated in Section III-D.
    \item \textbf{The application plane}: With massive data aggregated on this plane, many advanced applications can be implemented, such as IoT and metaverse.  
\end{itemize}

As aforementioned, SDN controllers bridge these planes and play a major role in managing the core wireless network. 
\textit{To be specific, the tasks of wireless network management can be summarized as follows.}
i) Defining the network routing and topology, thereby ensuring the connectivity among devices,
ii) allocating appropriate volumes of resources to deploy the slices for addressing certain computational, sensing, and communication requirements,
iii) exposing the available services to the application plane, allowing service providers to sell various network services (e.g., the rendering services in Fig. 2) directly to clients,
iv) maintaining the network, including authentication, network equipment monitoring, traffic/workload prediction, and anomaly detection.

\subsection{Major Issues in Wireless Network Management}
Multiple SDN controllers can be regarded as an integrated entity, called SDN orchestrator (see Fig. 2, Part B).
To handle the aforementioned tasks, the SDN orchestrator is typically equipped with various tools, such as contract theory \cite{contract}, stochastic optimization, and Deep Reinforcement Learning (DRL) models.
For instance, as a network economic approach, the contract theory helps design fair contracts between clients and service providers regarding provided service and the reward, which is significant for incentivizing the participation and improving the operational quality \cite{contract}.
%Likewise, QoS-oriented methods can measure non-functional characteristics (e.g., energy cost, latency, reliability) of atomic services, thereby guiding congestion control strategy and guarantying the service quality, especially for high-priority applications, with limited capacity.
For decision-making problems in wireless networks (e.g., resource allocation), DRL is regarded as one of the most efficient solutions \cite{routing}, \cite{Slice}.
Using a Markov decision process, DRL models enable the agents to learn the optimal \textit{action} under the given \textit{state} by setting different \textit{rewards}.
These techniques build the current paradigm for wireless network management.
Despite its wide adoption and mutuality, it suffers from the following issues.

\subsubsection{Training Data Scarcity}
Learning-based network management approaches typically require sufficient data for training the models. 
However, the wireless networks show great heterogeneity. 
Some slices might be idle due to the limited number of clients or the specific application types \cite{ToNTraffic}.
The third-party datasets containing real-world wireless network data also remain scarcely available and limited in size \cite{resource-2}.
Without enough training data, the models can hardly fully optimize their decision-making policies, affecting the efficiency of the wireless network management.

\subsubsection{Limited Flexibility}
Traditionally, the wireless network management is conducted by mapping a decision to a given state. 
% For example, contract theory generates contracts ($a$, $r$) according to the states of the service market. 
% Factors $a$ and $r$ indicate the expected service outcome and the reward for the service provider if the target can be achieved as promised, respectively. 
For example, the DRL models map each action in the current state to a reward and find the optimal actions leading to the maximum total reward. 
However, these methods fail to adapt to the randomness in wireless channels, caused by noise, multipath propagation, and interferences. 
For instance, Hua \textit{et al.} \cite{Slice} point out that given the random noise, each action of DRL corresponds to a reward distribution rather than a fixed reward value. 
Consequently, the generated management strategies can hardly reach the optimum.

\subsubsection{Dynamic Network States}
The structure of the wireless network states changes frequently over time due to the mobility of wireless devices. 
For instance, the initial \textit{Slice 3} in Fig. 2 contains three vehicles stopping at the cross and two pedestrians crossing the street.
A few minutes later, the pedestrians have left, and the vehicles have also moved their positions. 
The structure of network states, i.e., the number and connectivity of devices, changes accordingly. 
Traditional approaches, e.g., DRL, cannot well fit such mobility \cite{routing}. 
Recall that DRL models are trained to select the optimal action in the given state. 
If the state structure dramatically changes, we need to re-train the entire DRL model from scratch.

\subsection{Solutions from DGMs}
DGMs can address the above issues and further improve the efficiency of wireless network management, because of the following features.
\begin{itemize}
    \item \textbf{Generation ability}: DGMs can synthesize realistic content. Hence, we can realize the data augmentation using a small amount of real-world data and overcome the training data scarcity, which is one of the biggest constraints in wireless network management. 
    \item \textbf{Representation ability}: DGMs excel in representation, i.e., capturing semantic features embedded in the given samples via low-dimension latent vectors. Semantically similar samples can be mapped to similar areas in the latent space, even if their values are different (see Fig. 2). Here, samples can be modelled from the factors that we intend to explore in the wireless networks, e.g., resources and noise. Although directly interpreting the latent representations is difficult, DGMs can be trained to learn their distributions. With the distributions of the target factors, we can update our models/approaches accordingly, thereby improving their flexibility. Moreover, given that the states of the same network at different times follow the same pattern, i.e., are semantically similar, we can unify the latent spaces of the current network to the original one for overcoming the mobility. 
\end{itemize}

\subsection{DGM-Empowered Efficient Wireless Network Management}
As shown in Fig. 2, our framework accommodates multiple DGMs (including VAE, GAN, diffusion, etc.) into a module.
Acting as a flexible toolkit library, this module not only enables SDN orchestrators to undertake new tasks, but also helps them refine the existing approaches.
Assisted by DGMs, the workflows for managing wireless networks consist of the following four steps (Steps \circled{1}-\;\circled{4} in Fig. 2).
\begin{itemize}
    \item \textbf{Step 1: Model the network states.} Efficient network management can be regarded as an optimization problem. To find the optimal strategy, the SDN orchestrator should first model the network states according to the specific task. Taking network routing as an example, the topology, connectivity, bandwidth, and traffic usage pattern should be taken into consideration (see Fig. 2, \textit{Slice 3}). Such states are formed by high-dimensional vectors \cite{Slice}. Moreover, the management objective, i.e., minimizing the communication latency, should also be formulated. 
    \item \textbf{Step 2: Learn the latent representations.} Instead of directly working on the states, the proposed framework leverages DGMs to process such states and output their latent representations. The functions of latent representations are twofold. Firstly, they compress the network states with lower dimensions. As shown in Fig. 2, the complex network routing states of \textit{Slice 3} are represented by 2D vectors. The reduced size is conducive for storage and analysis. Meanwhile, latent representations are important knowledge since they capture the most important features of the current network states.
    \item \textbf{Step 3: Generate the management strategies.} With latent representations, in this step, we can generate the management strategies for wireless networks. Note that the implementation of this step is flexible. Firstly, some simple tasks can be easily completed by analyzing the latent representations (e.g., anomaly detection) or doing the inference using the well-trained DGMs in Step 2 (e.g., traffic prediction and data synthesis). To solve more challenging problems, such as routing and resource allocation, we can combine DGMs with other techniques, including contract theory, DRL \cite{routing}, etc. In the next part, we use several cases to show how DGMs can be combined with other techniques for further improving the efficiency of the wireless network management.
    \item \textbf{Step 4: Transfer the learned knowledge.} When encountering environment changes, the DGMs facilitate the knowledge transfer between different domains, denoted by source $S$ and target $T$ \cite{routing}. Note that this step mainly applies for learning-based approaches, e.g., DRL. To be specific, DGMs can align the structure of $S$ with that of $T$ via latent representations. Thus, the structural changes of the network states can be eliminated. Given that the \textit{action} space is also unchanged (\textit{action} is determined by the specific task and unrelated to the environment), the pre-trained DRL model can be efficiently generalized to $T$. Meanwhile, we can use samples from $T$ to fine-tune the model, further improving its adaptability across domains. More details are shown in Section III-E. 
\end{itemize}

\subsection{Use Cases}
In this part, we discuss several representative use cases for wireless network management leveraging DGMs.
Here, we only consider complicated tasks requiring the collaboration between DGMs and other techniques.

\subsubsection{Network Routing}
To manage a wireless network, the most crucial issue is routing, which determines the connectivity and networking performance. 
Nonetheless, routing is time-critical, requiring the SDN orchestrator to keep pace with the environment change and provide timely strategies.
Combining DGMs with DRL is proven to address this concern efficiently.
For instance, Dong \textit{et al.} \cite{routing} present a GAN-based DRL approach for rapidly transferring routing knowledge from $S$ to $T$.
To do so, a generator $\mathcal{G}_S$ is first trained to extract $R_S$, the latent representation of $S$.
Then, they train a DRL model, learning the routing strategies based on $R_S$ rather than $S$.
If the state structure changes to $T$, another generator $\mathcal{G}_T$ is used for generating $R_T$.
To align $S$ and $T$, a discriminator $\mathcal{D}_T$ is employed to distinguish $R_T$ and $R_S$, guiding $\mathcal{G}_T$ to learn $R_S$'s distribution. 
Clearly, $\mathcal{G}_S$, $\mathcal{G}_T$, and $\mathcal{D}_T$ construct a GAN. 
Using $R_T$ as the new $state$ for the previous DRL model, the knowledge from $S$ can be efficiently generalized to $T$.

\subsubsection{Resource Allocation}
Resource allocation is another major concern in wireless network management.
DGMs can help conventional learning-based approaches, e.g., DRL, by synthesizing more training samples.
For example, Kasgari \textit{et al.} \cite{resource-2} present an experienced DRL approach, in which a GAN-based refiner is developed to pre-train the model.
Particularly, the generator is feed with a mixed input: a few real-world data and massive synthetic data.
Then, they utilize a discriminator to eliminate the bias existing in the synthetic data and make it approach the real data.
In this way, firstly, they efficiently augment the dataset which emulates the given wireless environment.
Moreover, by fine-tuning the input synthetic data, some extreme network conditions can be reproduced, providing the model with rich experience.  

%Another flaw of traditional DRL is the inflexibility towards the randomness in the wireless networks.
%For instance, due to the random noise, each action in DRL actually corresponds to a reward distribution \cite{Slice}.
%Using generative models, e.g., VAE and GAN, such a distribution can be efficiently represented and learned.
%Consequently, the policies generated by DRL can become more flexible and robust. 
%Note that these proposals also follow the four steps defined in Section III-D.
%Due to the page limitation, we do not elaborate them step by step.

\subsubsection{Network Economics}
%From the user perspective, wireless networks should satisfy their service requests with high QoS.
%To do so, the SDN orchestrator relays users to the suitable service providers, which own enough resources and knowledge tailored to deal with the specific types of requests.
Network economics plays an important role in wireless networks for ensuring the participation of service providers \cite{contract}.
Nowadays, the mainstream incentive techniques include contract theory, auction theory, and game theory \cite{incentiveMechanism}.
Similar to routing and resource allocation, DGMs can be applied in these methods to achieve higher flexibility.
\textit{However, there still lacks the research using DGMs for designing incentive mechanisms in wireless networks.}
To this end, we conduct a case study, where we leverage the SOTA DGM, i.e., diffusion model, to generate contracts for incentivizing mobile AIGC services.
The methodology, implementation, and numerical results are demonstrated below.

\section{Case Study: Diffusion-empowered Contract Theory for Incentivizing Mobile AIGC Services}
In this part, we conduct a case study, presenting a diffusion-empowered contract theory for incentivizing mobile AIGC services. 
Our case study can fulfill the aforementioned research gap and further prove the effectiveness of DGMs in wireless network management.

\subsection{Background}
From the end of 2022, AIGC achieved phenomenal success, realizing the automatic content generation by machines \cite{yinqiu}.
Although such a paradigm greatly enriches the content diversity and saves labor, creating multimodal content from scratch is time-consuming and resource-intensive \cite{hongyang}.
Taking \textit{Stable Diffusion}, the SOTA text-to-image AIGC model, as an example, its inference on a single NVIDIA A100 GPU consumes 6.49 seconds and 7.69 GB memory (\textit{https://lambdalabs.com/blog/inference-benchmark-stable-diffusion}).
Obviously, such resource requirements are unaffordable for common clients.

To tackle this issue, some crowdsourcing platforms for AIGC services are proposed, aiming at exploiting the idle resources.
For instance, \textit{Stable Horde} accommodates hundreds of volunteers who contribute their idle computation resources and conduct AIGC inference for clients in need (\textit{https://aqualxx.github.io/stable-ui/}).
However, these volunteers currently work for free, which is unsustainable in the long term.
To this end, we intend to present an incentive mechanism for rewarding the AIGC service providers (ASPs) and thus encouraging their participation.
\begin{figure}[tpb]
\centering
\includegraphics[width=1\columnwidth]{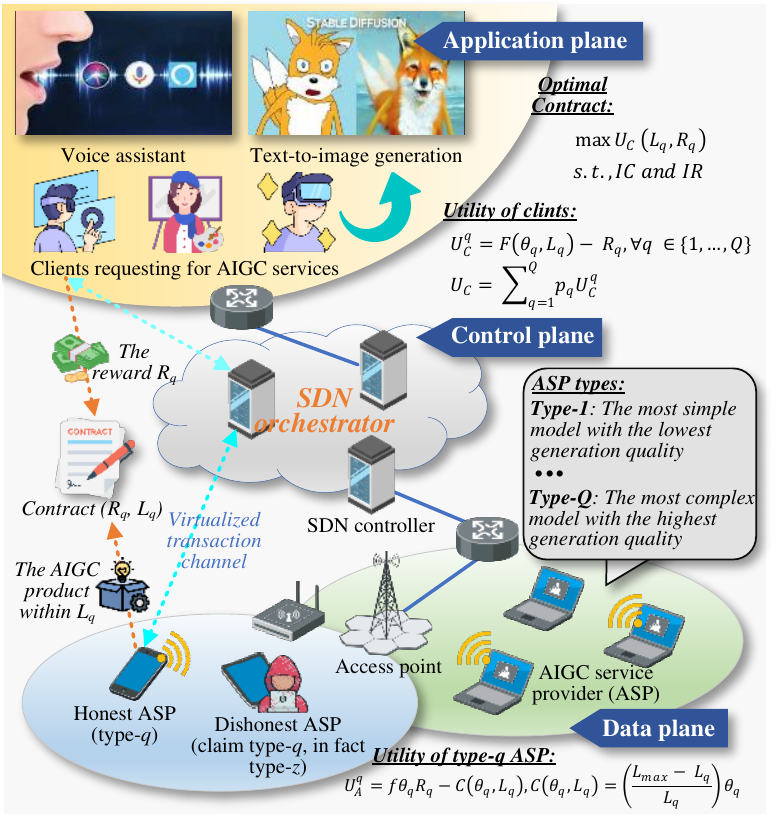}
\caption{The illustration of diffusion-empowered contract theory for incentivizing mobile AIGC services.}
\vspace{-0.5cm}
\label{ledger architecture}
\end{figure}

\subsection{System Model}
Fig. 3 illustrates the system model, which depicts a specific case of the proposed framework for supporting mobile AIGC.
Following the three-layer architecture, firstly, the data plane consists of numerous mobile devices (e.g., smartphones and laptops).
With certain computation and storage resources, they act as ASPs, which preserve well-trained AIGC models and conduct AIGC inferences for clients. 
The SDN orchestrator in the control plane manages the entire wireless network.
Particularly, to satisfy the massive requests for AIGC services, it constructs virtual links between ASPs and clients, allowing clients to directly make transactions with ASPs.
The clients are from the application plane, requiring AIGC services for realizing text-to-image generation, voice assistant, etc.

We propose a diffusion-based contract theory to incentivize ASPs.
Specifically, we suppose there exist $n$ ASPs in total.
According to the complexity of their local AIGC models (denoted by $\theta$), we split all ASPs into $Q$ types, satisfying $\theta_1 \leq \dots \leq \theta_Q$.
The higher the model complexity, the larger the model size and the better the inference quality (see Fig. 3).
The proportion of type-$q$ ASP in the network is $p_q$, $\forall q \in \{1, \dots, Q\}$.
In the service market, ASPs may publish their types to attract the clients. 
The clients, however, can hardly believe the information from ASPs in such an anonymous environment.
Contract theory can effectively address such information asymmetry by designing the specific contracts for every type of ASPs.
For type-$q$ ASP, it can only maximize its revenue by signing the type-$q$ contract.

\subsection{Problem Formulation}
Suppose that client $C$ intends to request for AIGC service.
To do so, it creates a set of contracts and waits for the ASPs to sign (see Fig. 3).
The form of contracts is \{$L_{q}$, $R_{q}$\}, in which $L_q$ and $R_q$ denote the latency requirement for and the reward to type-$q$ ASP, respectively.
The physical meaning of the contracts is that type-$q$ ASP can only receive $R_q$ by sending the AIGC products within $L_q$.

\subsubsection{Utility of Client}
The utility of $C$ towards type-$q$ ASP is denoted by $U_{C}^q$, equaling \textit{the revenue for receiving AIGC services within $L_q$} \texttt{minus} \textit{$R_q$}.
Note that the revenue, denoted by $F(\theta_q, L_q)$, is defined as a general security-latency metric, i.e, $e_1 (\theta_q)^{z_1} - e_2 (L_q/L_{max})^{z_2}$ \cite{reputationcontract}.
We can observe that this function is increasing w.r.t. $\theta_q$ and decreasing w.r.t. $L_q$.
These features indicate that $C$ prefers the ASPs with complicated models and providing AIGC services timely.
The objective of $C$ is to maximize $U_{C}$, by finding the optimal $L$ and $R$ for all types of ASPs.
The detailed equations regarding the clients' utility are shown in Fig. 3.

\subsubsection{Utility of ASP}
The utility of type-$q$ ASP is denoted by $U_A^q$, equaling the \textit{the evaluation toward the received reward} \texttt{minus} \textit{the invested resources}.
For simplicity, we formulate the resources invested by type-$q$ ASP, denoted by $c\,(\theta_q, L_q)$, based on the model complexity and the required latency (see Fig. 3).
Note that $L_{max}$ means the maximum expected service latency. 
Similar to $C$, ASPs also intend to maximize their utility.
To do so, they may hide their real type and sign the contracts targeting to other types of ASPs.

\subsubsection{Problem Formulation}
Given that contracts are generated by $C$, the overall optimization problem is: maximizing $U_{C}$.
Furthermore, to ensure the participation of ASPs, two constraints should be satisfied, namely Individual Rationality (IR) and Incentive Compatibility (IC) \cite{contract}.
IR means every ASP, no matter which type it belongs to, can earn non-zero utility by participating in this game.
Additionally, to eliminate the information asymmetry and prevent ASPs from cheating clients, IC guarantees that for arbitrary type-$q$ ASP, it can only receive the maximum utility when choosing the type-$q$ contract.
Then, the problem becomes finding the optimal \{$L_1^{\ast}, \dots, L_Q^{\ast}$\} and \{$R_1^{\ast}, \dots, R_Q^{\ast}$\}, which lead to the maximum $U_C$ while satisfying IR and IC.

\subsection{Diffusion-Empowered Contract Generation}
 % and Soft Actor-Critic (SAC)
Currently, researchers use DRL models, such as Proximal Policy Optimization (PPO), to find the optimal contracts. 
Inspired by diffusion Q-learning \cite{diffusioncontract}, we integrate the diffusion process in conventional DRL to improve its flexibility and exploration ability.
To help readers understand the four-step workflow mentioned in Section III-D, we elaborate our diffusion-empowered contract generation step-by-step.
Note that since this case study does not involve the knowledge transfer, only three steps are demonstrated.
\begin{itemize}
    \item \textbf{Step 1: Model the AIGC service market.} The state of the AIGC service market is formed by the vector $S$ := [$n$, $Q$, $L_{max}$, ($p_1, \dots, p_Q$), $(\theta_1, \dots, \theta_Q)$].
    \item \textbf{Step 2: Explore the policy in latent space.} We use a conditional diffusion model to exploit the latent space of contract generation \cite{diffusioncontract}. Specifically, we need to learn the policy $\pi(c^0|s)$ for generating the optimal contract $c^0$ under the state $s$ $\in \{S\}$. To do so, in each round, we start with a Gaussian noise $c^T$ and conduct denoising for $T$ times to acquire $c$. Each denoising step can be represented by $p(c^{i}|c^{i+1}, s)$ ($\forall i\in \{0, \dots, T-1\}$), which follows a normal distribution with learnable $mean$ and $variance$, as mentioned in Section II-D. $c^0$, $\dots$, $c^T$ are the latent space. The denoising in latent space can not only fine-tune the contract representation, but also facilitate the generator to explore more possibilities.
    \item \textbf{Step 3: Generate the contracts.} To train $mean$ and $variance$, we adopt an action-value function $Q(c^0|s)$, which is a Bellman combination of multiple $U_c$ under successive states. Then, we build a DRL model to gradually refine $\pi(c^0|s)$ until it can find the $c^0$ which leads to the highest $Q$. Specifically, we employ a double-Q DRL architecture \cite{diffusioncontract}, consisting of two Q-networks and two target networks. Two Q-networks are updated in an alternating fashion, i.e., each one learns from the other's experience to avoid overestimation. %The target networks are copies of the Q-network, while their weights are frozen (updated periodically) for stabilizing the training process. 
    The training objective is to minimize the overall difference between each Q-network and its target. In this way, we can acquire a precise a $Q$ function. Then, the $Q$ function can guide the diffusion process by evaluating each $c^0$. Finally, the optimal $c^0$ can be acquired.
\end{itemize}

\subsection{Numerical Results}
We conduct simulations to prove the validity of the proposed diffusion-empowered incentive mechanism.
Specifically, we suppose there exist 50 ASPs, divided into two types, i.e., $n$ = 50 and $Q$ = 2. 
$p_1$, $p_2$, and $L_{max}$ are generated randomly.
According to the model complexity (e.g., the number of parameters), $\theta_1$ and $\theta_2$ are randomly sampled within (10, 50) and (50, 100), respectively.
Weighting factors $f$, $e_1$, $e_2$, $z_1$, and $z_2$ are set to 0.05, 30, 5, 1, and 1, respectively.
%All the involved contract generators are trained on an Apple MacBook Pro with an 8-Core Intel Core i9 CPU and AMD Radeon Pro 5500M GPU.
Finally, the step number of the denoising processes is eight.
\begin{figure}[tpb]
\centering
\includegraphics[width=0.8\columnwidth]{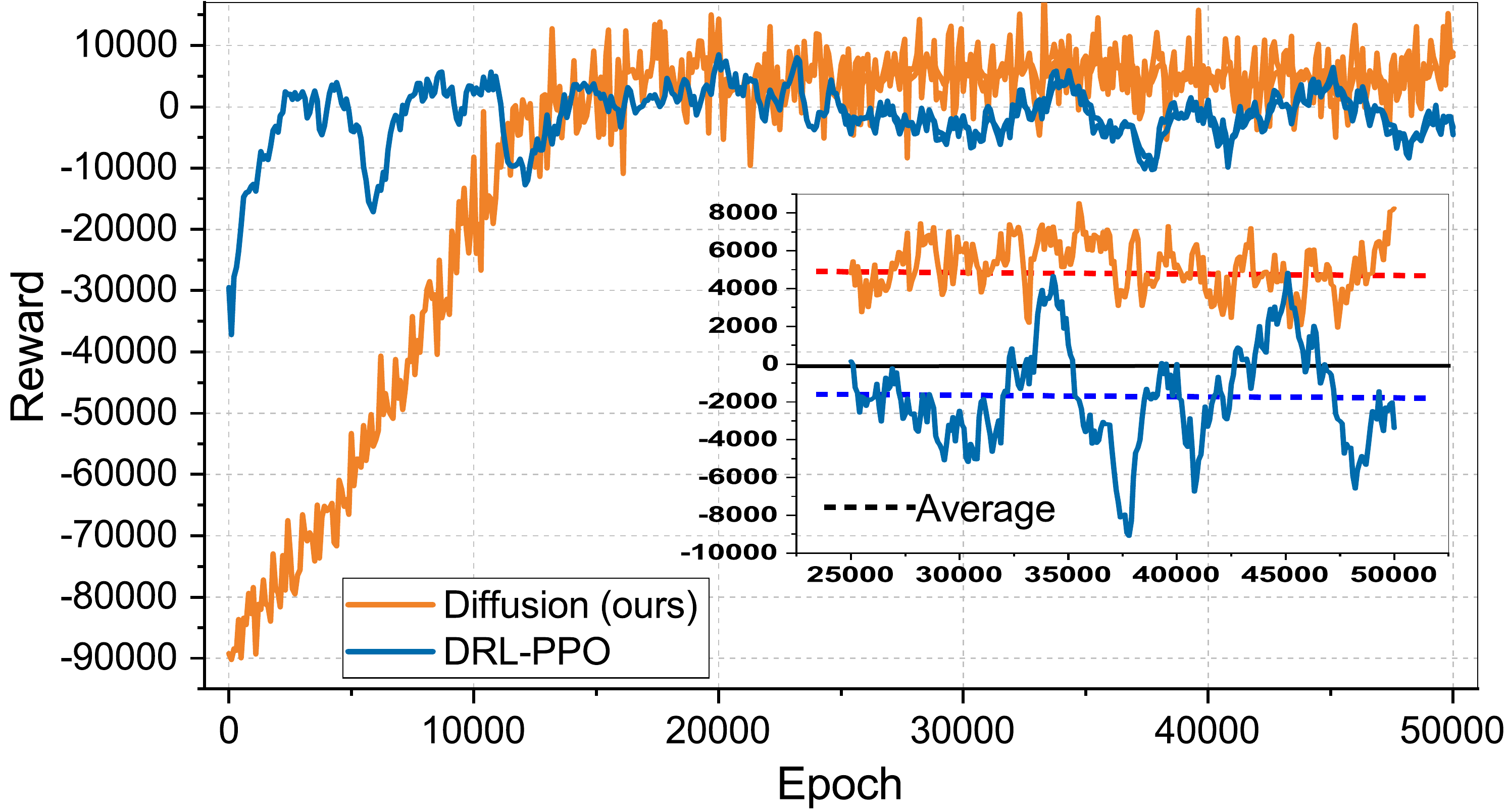}
\caption{Training curves of our diffusion-empowered approach and conventional PPO.}
\vspace{-0.3cm}
\label{ledger architecture}
\end{figure}
\begin{figure}[tpb]
\centering
\includegraphics[width=0.8\columnwidth]{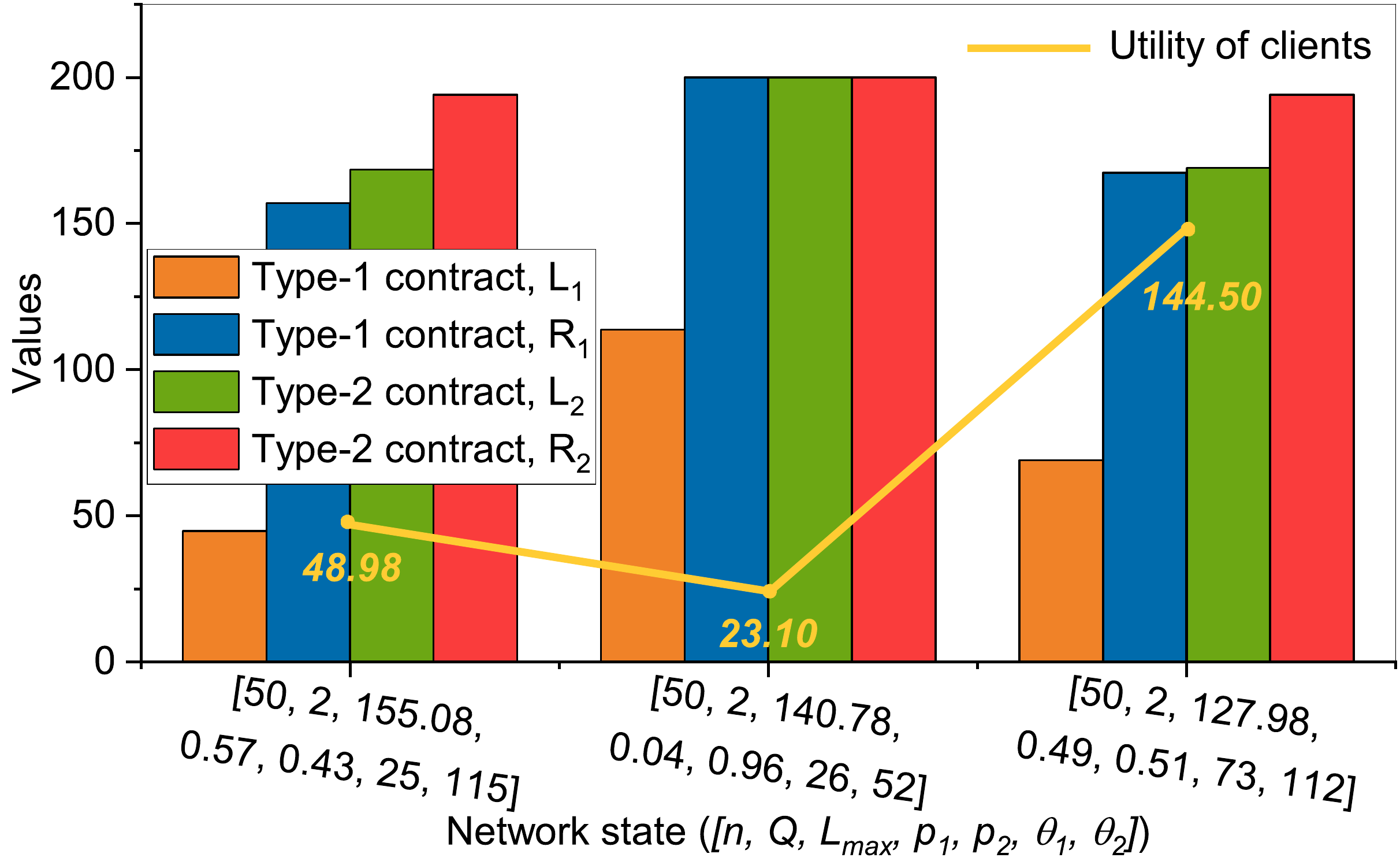}
\caption{The generated contracts under different states.}
\vspace{-0.5cm}
\label{ledger architecture}
\end{figure}

Fig. 4 depicts the training progress of the contract generators based on diffusion and conventional PPO. 
Although the diffusion method begins with a lower initial reward value, it can converge at about the same time as the PPO method. 
Notably, the reward values attained through the diffusion method are significantly higher than those obtained through PPO. 
Moreover, our approach realizes the stable contract generation, enabling clients to always obtain positive utilities (see the partially enlarged part of Fig. 4).
The superiority of the diffusion-empowered approach is mainly attributed to two reasons: i) the fine-grained policy tuning during the diffusion process can alleviate the effects caused by randomness and noise; ii) the exploration made by diffusion can improve the flexibility of the policies and avoid the model from falling into suboptimal solutions.
In contrast, the reward of PPO fluctuates around zero, which means the generated contracts easily fail to meet the IR or IC constraints.

Moreover, we inspect the quality of the contracts gendered by the diffusion-empowered approach.
Fig. 5 shows the details of the generated type-1 and type-2 contracts under different network states.
We can observe that contributed to the exploration experience during the denoising, our approach maintains high generation quality.
Following the contracts, the clients can acquire high rewards even in some extreme cases (e.g., the portion of $q_1$ is greatly low).

\section{Future Research Directions}

%\subsection{Interpretability of Latent Space Representations}
%Although latent space can embed rich semantic information, it is difficult to interpret.
%In this case, we can hardly capture the semantic meaning of the latent representations, even though they fit the real pattern with high fidelity.
%Such an issue prevents us from further improving the model generalization ability and reducing the latent space dimensions. 
%To this end, the interpretation methods for DGMs are worth research.

\subsection{Distributed and Energy-Efficient DGMs}
DGMs, especially pre-trained models, suffer from huge training costs, since learning complex distributions is tough.
To exploit the resources in wireless networks, the DGMs supporting coordinated and distributed training is worth studying.
Moreover, the inference of DGM is also resource-intensive, especially in wireless networks with limited power.
Therefore, how to incorporate the running costs of DGMs in sustainable network operation is also a crucial research topic.

\subsection{DGM-Aided Reconfigurable Intelligent Surface}
The wireless propagation conditions are becoming tougher due to the shortage of the available links. 
Reconfigurable Intelligent Surface (RIS) can provide supplementary links to improve the propagation environments.
However, the optimal beamforming design is a challenging issue because of the non-convex constraints on the RIS.
Similar to our case study, diffusion models show great potential to support current approaches, e.g., DRL, to achieve high utility.

\subsection{Governance Platforms for DGMs}
The strong capability of DGMs not only facilities the wireless network management, but also causes some cybersecurity concerns. 
The most famous one is \textit{deepfake}, in which attackers maliciously synthesize and distribute fake content (e.g., news and videos).
Moreover, wireless channels are vulnerable to eavesdropping, increasing the data leakage risk of DGMs.
%Furthermore, since DGMs can mimic the patterns of the given victims, they also help attackers generate countermeasures. 
%For example, MalGAN (\textit{https://github.com/lzylucy/Malware-GAN-attack}) can generate the malware which effectively avoids the detector, even though the detection rules are a black-box.
To govern and protect DGMs, the blockchain-based platforms and mechanisms are worth studying.
% Using an immutable blockchain ledger, the model operation and data usage can be protected.

\section{Conclusion}
This article conducts a systematic tutorial about why and how DGMs can improve the efficiency of wireless network management.
Specifically, we first introduce the basics of generative AI, several famous DGMs, and their applications.
Then, we present a general framework, on which we elaborate the steps for applying DGMs in wireless network management and show three use cases.
Furthermore, a case study is conducted, leveraging diffusion models to design flexible contracts for incentivizing mobile AIGC services.
Finally, we discuss several critical research directions which can further facilitate the application of DGMs in wireless network management. 
We hope that this article can inspire researchers to explore more possibilities of generative AI, as well as expanding its applications to more fields.
\vspace{0.3cm}

%\bibliography{aigcblockchain}

\bibliographystyle{IEEEtran}

\end{document}